\documentclass{easychair}

\usepackage{xspace}
\usepackage{xcolor}
\usepackage{tikz}

\newcommand{\codeintext}[1]{\texttt{#1}}

\usepackage{listings, xcolor}

\definecolor{verylightgray}{rgb}{.97,.97,.97}
\definecolor{darkgreen}{rgb}{0,.5,0}

\lstdefinelanguage{Solidity}{
  alsoletter={.},
  basicstyle=\scriptsize\ttfamily, 
	keywords=[1]{anonymous, assembly, assert, balance, break, call, callcode, case, catch, class, constant, continue, contract, debugger, default, delegatecall, delete, do, else, emit, event, export, external, false, finally, for, function, gas, if, implements, import, in, indexed, instanceof, interface, internal, is, length, library, log0, log1, log2, log3, log4, memory, modifier, new, payable, pragma, private, protected, public, pure, push, require, return, returns, revert, selfdestruct, send, storage, struct, suicide, super, switch, then, this, throw, transfer, true, try, typeof, using, view, while, with, addmod, ecrecover, keccak256, mulmod, ripemd160, sha256, sha3}, 
	keywordstyle=[1]\color{blue}\bfseries,
	keywords=[2]{address, bool, byte, bytes, bytes1, bytes2, bytes3, bytes4, bytes5, bytes6, bytes7, bytes8, bytes9, bytes10, bytes11, bytes12, bytes13, bytes14, bytes15, bytes16, bytes17, bytes18, bytes19, bytes20, bytes21, bytes22, bytes23, bytes24, bytes25, bytes26, bytes27, bytes28, bytes29, bytes30, bytes31, bytes32, enum, int, int8, int16, int24, int32, int40, int48, int56, int64, int72, int80, int88, int96, int104, int112, int120, int128, int136, int144, int152, int160, int168, int176, int184, int192, int200, int208, int216, int224, int232, int240, int248, int256, mapping, string, uint, uint8, uint16, uint24, uint32, uint40, uint48, uint56, uint64, uint72, uint80, uint88, uint96, uint104, uint112, uint120, uint128, uint136, uint144, uint152, uint160, uint168, uint176, uint184, uint192, uint200, uint208, uint216, uint224, uint232, uint240, uint248, uint256, var, void, ether, finney, szabo, wei, days, hours, minutes, seconds, weeks, years},	
	keywordstyle=[2]\color{teal}\bfseries,
	keywords=[3]{block, blockhash, coinbase, difficulty, gaslimit, number, timestamp, msg.sender, msg.value, gas, sig, now, tx, gasprice, origin},	
	keywordstyle=[3]\color{violet}\bfseries,
	identifierstyle=\color{black},
	sensitive=true,
	comment=[l]{//},
	morecomment=[s]{/*}{*/},
	commentstyle=\color{darkgreen}\ttfamily,
	stringstyle=\color{red}\ttfamily,
	morestring=[b]',
	morestring=[b]",
	numberstyle=\scriptsize,
	emphstyle=\underline
}

\lstdefinelanguage{Boogie}{
  basicstyle=\scriptsize\ttfamily, 
	keywords=[1]{axiom, break, call, const, else, exists, extends, forall, function, goto, if, implementation, modifies, old, procedure, returns, then, type, unique, var, while}, 
	keywordstyle=[1]\color{blue}\bfseries,
	keywords=[2]{address, bool, int, real}, 
	keywordstyle=[2]\color{teal}\bfseries,
	keywords=[3]{assert, assume, ensures, invariant, requires}, 
	keywordstyle=[3]\color{darkgreen}\bfseries,
	keywords=[4]{\_msg\_sender, \_msg\_value, \_balance, \_this}, 
	keywordstyle=[4]\color{violet}\bfseries,
	identifierstyle=\color{black},
	sensitive=false,
	comment=[l]{//},
	morecomment=[s]{/*}{*/},
	commentstyle=\color{gray}\ttfamily,
	stringstyle=\color{red}\ttfamily,
	morestring=[b]',
	morestring=[b]",
  numberstyle=\scriptsize
}

\newcommand{\solcverify}{\textsc{solc-verify}\xspace}

\newcommand{\trackschangesin}{\codeintext{tracks-changes-in}\xspace}

\title{Formal Specification and Verification of Solidity Contracts with Events}

\author{
	\'{A}kos Hajdu\inst{1}%
	\and
	Dejan Jovanovi\'{c}\inst{2}%
	\and
	Gabriela Ciocarlie\inst{2}%
}

\institute{
	Budapest University of Technology and Economics, Budapest, Hungary\\
	\email{hajdua@mit.bme.hu}
	\and
	SRI International, New York City, USA\\
	\email{\{dejan.jovanovic,gabriela.ciocarlie\}@sri.com}
}

\authorrunning{\'A. Hajdu, D. Jovanovi\'{c} and G. Ciocarlie}

\titlerunning{Formal Spec. and Verif. of Solidity Contracts with Events}

\begin{document}

\maketitle

\begin{abstract}
Events in the Solidity language provide a means of communication between the on-chain services of decentralized applications and the users of those services.
Events are commonly used as an abstraction of contract execution that is relevant from the users' perspective.
Users must, therefore, be able to understand the meaning and trust the validity of the emitted events.
This paper presents a source-level approach for the formal specification and verification of Solidity contracts with the primary focus on events.
Our approach allows specification of events in terms of the on-chain data that they track, and predicates that define the correspondence between the blockchain state and the abstract view provided by the events.
The approach is implemented in \solcverify, a modular verifier for Solidity, and we demonstrate its applicability with various examples.
\end{abstract}




\section{Introduction}

Ethereum is a public, blockchain-based computing platform supporting the development of decentralized applications~\cite{wood2014ethereum}.
The core of such applications are programs -- termed smart contracts~\cite{szabo1994smart} -- deployed on the blockchain.
While Ethereum nodes run a low-level virtual machine (EVM~\cite{wood2014ethereum}), smart contracts are usually written in a high-level, contract-oriented language, most notably Solidity~\cite{soliditydoc}.
The contract code can be executed by issuing transactions to the network, which are then processed by the participating nodes.
Results of a completed transaction are provided to the issuing user, and other interested parties observing the contract, through transaction receipts. While the blockchain is publicly available for users to inspect and replay the transactions, the contracts can communicate important state changes, including intermediate changes, by emitting events~\cite{consensys2016guide}.
Events usually represent a limited abstract view of the transaction execution that is relevant for the users, and they can be read off the transaction receipts.
The common expectation is that by observing the events, the user can reconstruct the relevant parts of the current state of the contracts.
Technically, events can be viewed as special triggers with arguments that are stored in the blockchain logs.
While these logs are programmatically inaccessible from contracts, the users can easily subscribe to and observe the events with the accompanying data.
For example, a token exchange application can monitor the current state of token balances by tracking transfer events in the individual token contracts.

Smart contracts, as any software, are also prone to bugs and errors.
In the Ethereum context, any flaws in contracts come with potentially devastating financial consequences, as demonstrated by various infamous examples~\cite{atzei2017survey}.
While there has been a great interest in applying formal methods to smart contracts~\cite{atzei2017survey,chen2019survey}, events are usually considered merely a logging mechanism that is not relevant for functional correctness.
However, since events are a central state-change notification mechanism for users of decentralized applications, it is crucial that the users are able to understand the meaning and trust the validity of the emitted events.
In this paper, we propose a source-level approach for the formal specification and verification of Solidity contracts with the primary focus on events.
Our approach provides in-code annotations to specify events in terms of the blockchain data they track, and to declare events possibly emitted by functions.
We verify that (1) whenever tracked data changes, a corresponding event is emitted, and (2) an event can only be emitted if there was indeed a change.
Furthermore, to establish the correspondence between the abstract view provided by events and the actual execution, we allow events to be annotated with predicates (conditions) that must hold before or after the data change.
We implemented the proposed approach in the open-source%
\footnote{\url{https://github.com/SRI-CSL/solidity/tree/merge}}
\solcverify~\cite{vstte2019,esop2020} tool and demonstrated its applicability via various examples.
\solcverify is based on modular program verification, but we present our idea in a more general setting that can serve as a building block for alternative verification approaches.

\section{Background}

Solidity~\cite{soliditydoc} is a high-level, contract-oriented programming language supporting the rapid development of smart contracts for the Ethereum platform.
We briefly introduce Solidity by restricting our presentation to the aspects relevant for events.
An example contract (\codeintext{Registry}) is shown in Figure~\ref{fig:registry}.
Contracts are similar to classes in object-oriented programming.
A contract can define additional types, such as the \codeintext{Entry} struct in the example, consisting of a Boolean flag and an integer data.
The persistent data stored on the blockchain can be defined with \emph{state variables}.
The example contract declares a single variable \codeintext{entries}, which is a mapping from addresses
to \codeintext{Entry} structs.
Contracts can also define \emph{events} including possible arguments.
The example declares two events, \codeintext{new\_entry} and \codeintext{updated\_entry}, to signal a new or an updated entry, respectively.
Both events take the address and the new value for the data as their arguments.
Finally, functions are defined that can be called as transactions to act on the contract state.
The example defines two functions: \codeintext{add} and \codeintext{update}.
The \codeintext{add} function first checks with a \codeintext{require} that the data corresponding to the caller address (\codeintext{msg.sender}) is not yet set.
If the condition of \codeintext{require} does not hold, the transaction is reverted. Otherwise, the function sets the data and the flag, and emits the \codeintext{new\_entry} event.
The \codeintext{update} function is similar to \codeintext{add}, with the exception that the data must already be set, and the new value should be larger than the old one (for illustrative purposes).

Note that Solidity puts no restrictions on the emitted events, and a faulty (or malicious) contract could both emit events that do not correspond to state changes or miss triggering an event on some change~\cite{chen2020defining}, potentially misleading users. In the case of the \codeintext{Registry} contract, the events are emitted correctly, and the user can reproduce the changes in \codeintext{entries} by relying solely on the emitted events and their arguments.

\begin{figure}[htb]
\begin{lstlisting}[language=Solidity,firstline=3]
contract Registry {
  struct Entry { bool set; int data; } // User-defined type

  mapping(address=>Entry) entries; // State variable

  /// @notice tracks-changes-in entries
  /// @notice precondition !entries[at].set
  /// @notice postcondition entries[at].set && entries[at].data == value
  event new_entry(address at, int value);

  /// @notice tracks-changes-in entries
  /// @notice precondition entries[at].set && entries[at].data < value
  /// @notice postcondition entries[at].set && entries[at].data == value
  event updated_entry(address at, int value);

  /// @notice emits new_entry
  function add(int value) public {
    require(!entries[msg.sender].set);
    entries[msg.sender].set = true;
    entries[msg.sender].data = value;
    emit new_entry(msg.sender, value);
  }

  /// @notice emits updated_entry
  function update(int value) public {
    require(entries[msg.sender].set && entries[msg.sender].data < value);
    entries[msg.sender].data = value;
    emit updated_entry(msg.sender, value);
  }
}
\end{lstlisting}
\caption{An example contract illustrating Solidity events. Users of the contract can associate an integer value to their address and can later update it with a larger value.}
\label{fig:registry}
\end{figure}

\solcverify~\cite{vstte2019} is a source-level verification tool for checking  functional correctness of smart contracts.
\solcverify takes contracts written in Solidity and provides various in-code annotations to specify functional behavior (e.g., pre- and postconditions, invariants).
\solcverify translates the annotated contracts to the Boogie Intermediate Verification Language (IVL) and uses the Boogie verifier~\cite{barnett2006boogie} to perform modular verification by discharging verification conditions to SMT solvers.
This paper presents extensions to the specification and translation capabilities of \solcverify that enable reasoning about Solidity events.
We propose event-specific annotations (Section~\ref{sec:eventspec}) and use them to instrument the code during translation with additional conditions to be verified (Section~\ref{sec:verification}).

\section{Specification of Events}
\label{sec:eventspec}

Our approach provides in-code annotations to specify events in terms of the on-chain data that they track for changes.
Furthermore, additional predicates can specify the correspondence between the abstract view provided by events and the actual data, before and after the change.

\paragraph{Data changes and checkpoints.}

Each event can declare a set of contract state variables that it \emph{tracks} for changes.
In the \codeintext{Registry} example (Figure~\ref{fig:registry}), both events track the single state variable \codeintext{entries}, as specified by the \trackschangesin annotations.
Intuitively, we use the tracking of changes to make sure that
(1) if a tracked variable changes, a corresponding event must be emitted after; and
(2) an event should be emitted only if some of its tracked variables have changed before.
As data changes often occur in multiple steps (e.g., updating both members of a struct in the function \codeintext{add} of Figure~\ref{fig:registry}), or conditionally, events cannot always be emitted directly after a single modifying statement.
Therefore, we define the precise semantics of ``before'' and ``after'' by introducing \emph{before-} and \emph{after-checkpoints}.
Before-checkpoints of an event are determined dynamically by the first change in a variable they track.
In contrast, after-checkpoints are defined by static barriers, marking the latest point in code where the emitting should be fulfilled.
Currently, we define loop and transaction boundaries (external calls to public functions and function return) as after-checkpoints.
The semantics of checkpoints is that an event corresponding to a state variable change must be emitted at some point between before- and after-checkpoints, which also clears the before-checkpoint. Conversely, an event can only be emitted if a tracked variable indeed changed (there was a before-checkpoint).

\paragraph{Event pre- and postconditions.}

In addition to the set of tracked variables, events can also be annotated with \emph{predicates} that define conditions over the state variables and the arguments of the event.
There are two kinds of predicates: \emph{pre-} and \emph{postconditions}.
Preconditions capture the values of state variables at the before-checkpoint, while postconditions correspond to the point when the event is emitted.
In the \codeintext{Registry} contract (Figure~\ref{fig:registry}), both events (\codeintext{new\_entry} and \codeintext{updated\_entry}) have the same postcondition, namely that the data at the given address must be set and its value must match the value in the argument.
The precondition of \codeintext{new\_entry} is that the data must not yet be set, while for \codeintext{updated\_entry}, it must be set and its value should be smaller than the event argument.
Postcondition expressions often need to connect the state at the point of emit and before the change.
As an example, consider the \codeintext{transfer} function of the token contract in Figure~\ref{fig:token} that deducts the sender's balance and increases the receiver's. To specify the postcondition of the \codeintext{Transfer} event, we need to relate the new balances to the previous balances.
We provide a special \codeintext{before} function -- to be used in postconditions -- that refers to previous values of state variables.

\paragraph{Functions.}

We require contract functions to be annotated with the events that they possibly emit using the \codeintext{emits} keyword.
For example, the \codeintext{add} and \codeintext{update} functions in Figure~\ref{fig:registry} can emit \codeintext{new\_entry} and \codeintext{updated\_entry} respectively.
If a function calls other functions (including base constructors), the callee's emitted events must also be included in the caller's specifications.

\begin{figure}[htb]
\begin{lstlisting}[language=Solidity,firstline=3]
contract Token {
  mapping(address=>uint) balances;

  /// @notice tracks-changes-in balances
  /// @notice precondition balances[from] >= amount
  /// @notice postcondition balances[from] == before(balances[from]) - amount
  /// @notice postcondition balances[to] == before(balances[to]) + amount
  event Transfer(address from, address to, uint amount);

  /// @notice emits Transfer
  function transfer(address to, uint amount) public {
    require(balances[msg.sender] >= amount && msg.sender != to);
    balances[msg.sender] -= amount;
    balances[to] += amount;
    emit Transfer(msg.sender, to, amount);
  }
}
\end{lstlisting}
	\caption{A token contract illustrating event postconditions that refer to previous state.}
	\label{fig:token}
\end{figure}

\section{Verification}
\label{sec:verification}

A contract with events and specifications is checked in two steps.
First, a syntactical check is performed to ensure that functions only emit events that they specified (via \codeintext{emits} annotations).
Then, we check the data tracking specifications and predicates by translating the contract to the input language of a verifier and instrumenting the code with the checks and the required bookkeeping.
In our implementation, we use the Boogie IVL and verifier~\cite{barnett2006boogie}, but we present our solution in a general way that can be reused in other Solidity verifiers.

\paragraph{Function emits.}

We first check whether functions only emit those events that are specified via \codeintext{emits} annotations.
This is a syntactic check on the Solidity AST: we find all emit statements in the function and check whether the corresponding events are specified to be emitted.
When a function calls other functions \emph{internally} (i.e., from the same contract), we apply a modular check based on the call graph: all events specified to be emitted by the callee must also be specified by the caller.
On the other hand, we currently ignore \emph{external} calls (such as \codeintext{.call()} or \codeintext{.transfer()}).
Such external calls cannot modify state variables or trigger events from the current contract directly (as they are non-public).
Indirect modifications and emits are possible by calling back public functions, but those are specified and checked independently (modularity of reasoning).
Finally, we also verify at each assignment (to a tracked variable), whether the function specifies a corresponding event to be emitted.

\paragraph{Data tracking and predicates.}

\begin{figure}[tb]
\begin{tikzpicture}
\tikzstyle{lbl}=[rectangle,inner sep=0.03cm,minimum height=0.32cm,fill=gray,text=white,font=\scriptsize\sffamily]

\node at (-4,0) {
\begin{lstlisting}[language=Solidity,linewidth=6.5cm]
contract C {
  uint x;
  uint y;

  /// @notice tracks-changes in x
  /// @notice tracks-changes in y
  /// @notice precondition x <= y
  /// @notice postcondition x <= y
  /// @notice postcondition x + diff == y
  event xy_changed(uint diff);

  /// @notice emits xy_changed
  function f(uint n) public {

    require(x <= y);

    y += n;

    emit xy_changed(y - x);

    for (uint i = 0; i < n; ++i) {

      x++;

      emit xy_changed(y - x);

    }

  }
}
\end{lstlisting}
};

\node (newvars) at (2,3.3) {
\begin{lstlisting}[language=Solidity,linewidth=8cm,numbers=none]
uint x_old; // Previous state of x
uint y_old; // Previous state of y
bool x_mod; // Modified since last checkpoint
bool y_mod; // Modified since last checkpoint
\end{lstlisting}
};
\node [lbl,minimum width=1.5cm] at (newvars.north west) [anchor=west] {new-vars};
\node [lbl] at (-6.9,3.22) [anchor=west] {new-vars};

\node (assumeclear) at (3.2,1.9) {
\begin{lstlisting}[language=Solidity,linewidth=6.6cm,numbers=none]
require(!x_mod && !y_mod); // Modif. clear
\end{lstlisting}
};
\node [lbl,minimum width=1.5cm] at (assumeclear.north west) [anchor=west] {assume-clear};
\node [lbl] at (-6.6,0.42) [anchor=west] {assume-clear};

\node (checkclear) at (3.12,0.9) {
\begin{lstlisting}[language=Solidity,linewidth=6.6cm,numbers=none]
assert(!x_mod && !y_mod); // Modif. clear
\end{lstlisting}
};
\node [lbl,minimum width=1.5cm] at (checkclear.north west) [anchor=west] {after-chpt};
\node [lbl] at (-6.6,-1.28) [anchor=west] {after-chpt};
\node [lbl] at (-6.3,-3) [anchor=west] {after-chpt};
\node [lbl] at (-6.6,-3.55) [anchor=west] {after-chpt};

\node (ymodif) at (3.2,-0.2) {
\begin{lstlisting}[language=Solidity,linewidth=6.6cm,numbers=none]
// Save y if not saved yet: before-checkpt
if (!y_mod) { y_old = y; y_mod = true; }
\end{lstlisting}
};
\node [lbl,minimum width=1.5cm] at (ymodif.north west) [anchor=west] {y-before};
\node [lbl] at (-6.6,-0.15) [anchor=west] {y-before};

\node(emitspec) at (2.4,-1.9) {
\begin{lstlisting}[language=Solidity,linewidth=8cm,numbers=none]
assert(x_mod || y_mod); // Emit without change
assert((x_mod?x_old:x) <= (y_mod?y_old:y)); // Pre
assert(x <= y); // Post
assert(x + (y - x) == y); // Post
x_mod = y_mod = false; // Emitted
\end{lstlisting}
};
\node [lbl,minimum width=1.5cm] at (emitspec.north west) [anchor=west] {emit-spec};
\node [lbl] at (-6.6,-0.74) [anchor=west] {emit-spec};
\node [lbl] at (-6.3,-2.4) [anchor=west] {emit-spec};

\node (xmodif) at (3.2,-3.6) {
\begin{lstlisting}[language=Solidity,linewidth=6.6cm,numbers=none]
// Save x if not saved yet: before-checkpt
if (!x_mod) { x_old = x; x_mod = true; }
\end{lstlisting}
};
\node [lbl,minimum width=1.5cm] at (xmodif.north west) [anchor=west] {x-before};
\node [lbl] at (-6.3,-1.83) [anchor=west] {x-before};

\end{tikzpicture}
\caption{Example contract with instrumentation snippets for checking event specifications.}
\label{fig:instrumentation}
\end{figure}
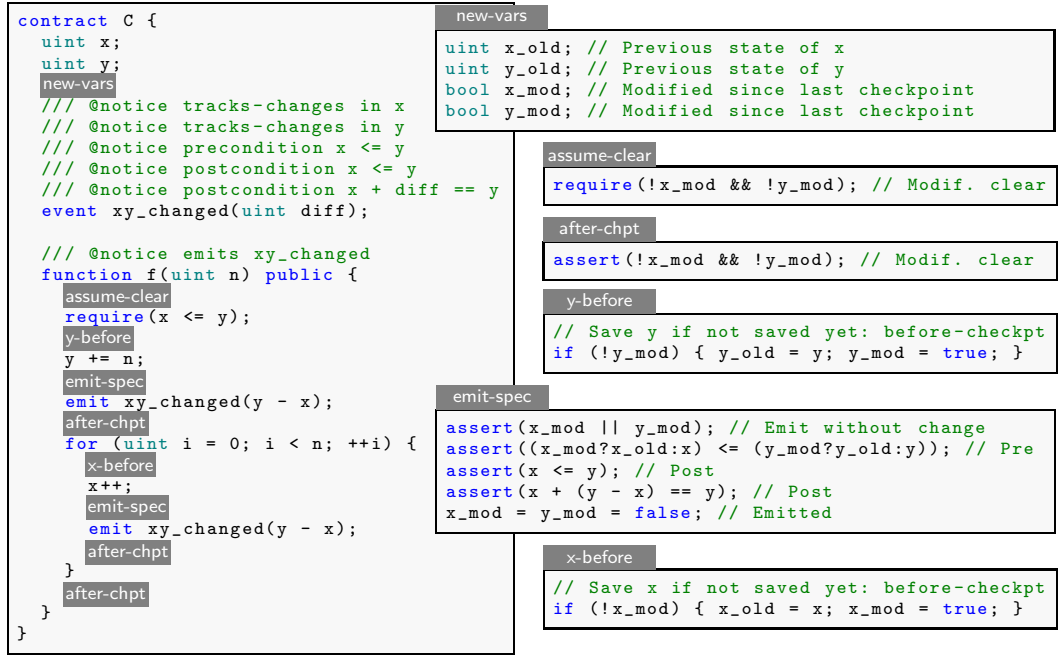

Verification of data tracking and predicates is performed by instrumenting the contract code with additional variables and statements to save state and to make extra checks at checkpoints.
For clarity, we describe the instrumentation on the Solidity level.
We illustrate the approach through the example contract in Figure~\ref{fig:instrumentation}, which has two state variables \codeintext{x} and \codeintext{y}, and whenever one of them changes, an event is emitted with their current difference.
Furthermore, \codeintext{x <= y} should hold both at the before- and the after-checkpoint.
The extra instructions are displayed as labels where they are injected, while the corresponding code can be found in the snippets to the right.

For each state variable that is tracked by any event, we introduce two additional variables in the contract: one with the same type to save the before-state, and a Boolean flag to indicate whether the data has been modified (snippet \textsf{new-vars} in Figure~\ref{fig:instrumentation}).
Functions are then instrumented with extra statements to save state, enforce after-checkpoints (barriers) and to perform specification checks when events are emitted.
Functions ensure on entry that none of the variables tracked by their specified events have been modified since the checkpoint before the call (snippet~\textsf{assume-clear}).
In other words, all relevant events must have been emitted before making the call.
In modular verification, this assumption becomes a precondition to the function.
Before each modification (assignment statement), if the state variable is not modified yet, the current value is stored%
\footnote{Saving data (e.g., mappings) with assignments might not yield valid Solidity code.
This code is for clarity of presentation and is handled by \solcverify  internally.}
in the helper variable and the flag for modification is set, introducing a before-checkpoint (snippets~\textsf{y-before} and \textsf{x-before}).

At each \codeintext{emit} statement, several checks are added (snippet~\textsf{emit-spec}).
First, we check that the data has indeed been modified, otherwise the event should not be emitted.
Then we check each pre- and postcondition.
By default, preconditions refer to the before-state and postconditions to the current values, except if the variable is explicitly wrapped with \codeintext{before()}.
Note that we refer to the previous value of a variable \codeintext{v} with \codeintext{v\_mod ?\ v\_old :\ v} because in general there might be variables that were not modified (e.g., \codeintext{x} at the first emit in Figure~\ref{fig:instrumentation}).
After performing the checks, emitting the event clears the flags (before-checkpoints).
Finally, before returning, functions enforce after-checkpoints by asserting that no state variable is in a modified state, i.e., the function cannot end in debt with events (snippet \textsf{after-chpt}).
In modular verification, this check becomes a postcondition to the function.
We also insert an after-checkpoint before the loop and at the end of every iteration (serving as loop invariant).

\paragraph{Discussion.}

One potential limitation of our approach is that we consider loop boundaries after-checkpoints: some contracts change the data many times in the loop but only emit a single summarizing event after the loop.
This limitation could be alleviated with annotations to ``allow delaying'' the emit after the loop, but we do not support this as it leads to more complex  specification and verification.

Our approach is not tied to Boogie or modular verification.
The instrumentation can be performed on the Solidity level, and the correctness of the specification is reduced to checking assertions at particular points in the code.
This means that the instrumented code can be fed into any Solidity verifier that can check for assertion failures.
The event specifications are deemed correct if and only if there are no related assertion failures.

A possible future use-case of our approach lies in the behavioral analysis of contracts based on logs.
Such analyses could reveal relationships individually and across contracts that are not otherwise apparent (e.g., exposing entities that control the blockchain interactions) or attack evidence.
Application-level log analysis has been used for a long time for monitoring and security purposes, and most existing techniques assume that application logs can be trusted or, if applications are subverted by attackers, the subversion can be captured~\cite{203706}. Our approach guarantees the validity of the emitted events, making them even more suitable for such analysis.


\section{Conclusion}

We presented an approach for the formal specification and verification of Solidity smart contracts that rely on events to communicate with their users, providing an abstract view of their state.
We proposed in-code annotations to specify events in terms of the state variables they track for changes.
Furthermore, we introduced additional predicates (pre- and postconditions) for specifying conditions on the state before and after the change, establishing the correspondence between the blockchain state and the emitted events.
The approach is implemented in \solcverify and we demonstrated its applicability with various examples.


\bibliographystyle{plain}
\bibliography{references}


\end{document}